\documentstyle[12pt,epsf]{article}
 \newcommand{\insertplot}[5]{\begin{figure}
 \hfill\hbox to 0.05in{\vbox to #5in{\vfill
 \inputplot{#1}{#4}{#5}}\hfill}
 \hfill\vspace{-.1in}
 \caption{#2}\label{#3}
 \end{figure}}
 \newcommand{\inputplot}[3]{
 \special{ps: plotfile #1}
}

\begin{document}
 
\title{Gravitating Dyons and Dyonic Black Holes}
\vspace{1.5truecm}
\author{
{\bf Yves Brihaye$^1$, Betti Hartmann$^2$,
%Burkhard Kleihaus$^3$,
and Jutta Kunz$^2$}\\
$^1$ Facult\'e des Sciences, Universit\'e de Mons-Hainaut\\
B-7000 Mons, Belgium\\
$^2$Fachbereich Physik, Universit\"at Oldenburg, Postfach 2503\\
D-26111 Oldenburg, Germany}
 
%\date{March 13, 1995}
 
\maketitle
\vspace{1.0truecm}

\begin{abstract}
We study static spherically symmetric gravitating dyon solutions
and dyonic black holes in Einstein-Yang-Mills-Higgs theory.
The gravitating dyon solutions share many features with the
gravitating monopole solutions.
In particular, gravitating dyon solutions and dyonic black holes exist 
only up to a maximal coupling constant,
and beside the fundamental dyon solutions
there are excited dyon solutions.
\end{abstract}
\vfill
\noindent {Preprint hep-th/9807169} \hfill\break
\vfill\eject

\section{Introduction}

SU(2) Einstein-Yang-Mills-Higgs (EYMH) theory, with the Higgs field
in the adjoint representation,
possesses globally regular gravitating magnetic monopole solutions
and corresponding magnetically charged black hole solutions
\cite{ewein,bfm,aichel}.
For small gravitational constant,
the gravitating fundamental monopole solution smoothly emerges
from the corresponding flat space solution,
the 't Hooft-Polyakov monopole \cite{thooft}.
With increasing gravitational constant,
the mass of the gravitating fundamental monopole solution decreases, and
it ceases to exist beyond a maximal value of the gravitational constant.

The corresponding magnetically charged EYMH black hole solutions
represent counterexamples to the ``no-hair'' conjecture.
Distinct from embedded Reissner-Nordstr\o m (RN) black holes
with unit magnetic charge,
they emerge from the globally regular magnetic monopole solutions 
when a finite regular event horizon is imposed. Consequently,
they have been characterized as ``black holes within magnetic monopoles''
\cite{ewein}.

Besides the fundamental monopole solution 
there are excited monopole solutions \cite{bfm}.
The gauge field function of the $n$-th excited monopole solution
possesses $n$ nodes,
whereas the gauge field function of the fundamental monopole solution 
decreases monotonically to zero \cite{bfm}.
The excited monopole solutions are related to the globally regular
Einstein-Yang-Mills (EYM) solutions, 
found by Bartnik and McKinnon \cite{bm},
and, like these solutions, have no flat space counterparts.

In flat space also dyon solutions exist, carrying both
electric and magnetic charge \cite{jul}.
Here we show that, like the monopole solutions,
these dyon solutions persist in the presence of
gravity, up to some maximal value of the gravitational constant.
Beside the fundamental gravitating dyon solutions,
we also construct excited gravitating dyon solutions
and dyonic black hole solutions.
Distinct from the embedded RN solutions
with the same electric and magnetic charge,
these ``black holes within dyons'' again represent counterexamples
to the ``no-hair'' conjecture.
In contrast, pure
SU(2) EYM theory possesses neither regular dyon solutions \cite{gal1},
nor dyonic black holes other than embedded RN solutions \cite{gal2,popp}.

\section{Einstein-Yang-Mills-Higgs Equations of Motion}

We consider the SU(2) EYMH action
\begin{equation}
S=S_G+S_M=\int L_G \sqrt{-g} d^4x + \int L_M \sqrt{-g} d^4x
\ \label{action}   \end{equation}
with
\begin{equation}
L_G=\frac{1}{16\pi G}R
\ , \end{equation}
and
\begin{equation}
L_M = - \frac{1}{4} F_{\mu\nu}^a F^{a\mu\nu} 
      - \frac{1}{2} D_\mu \phi^a D^\mu \phi^a
      - \frac{1}{4} \lambda (\phi^a \phi^a - v^2)^2
\ , \end{equation}
where
\begin{equation}
F_{\mu\nu}^a = \partial_\mu A_\nu^a - \partial_\nu A_\mu^a
 + e \epsilon^{abc} A_\mu^b A_\nu^c
\ , \end{equation}
\begin{equation}
D_\mu \phi^a = \partial_\mu \phi^a + e \epsilon^{abc} A_\mu^b \phi^c
\ , \end{equation}
$e$ is the gauge coupling constant, $\lambda$ is the Higgs coupling
constant and $v$ is the Higgs field vacuum expectation value.
Variation of the action eq.~(\ref{action}) with respect to the metric
$g_{\mu\nu}$, the gauge field $A_\mu^a$ and the Higgs field $\phi^a$
leads to the Einstein equations and the matter field equations.
 
To construct static spherically symmetric globally regular
and black hole solutions
we employ Schwarz\-schild-like coordinates and adopt
the spherically symmetric metric 
\begin{equation}
ds^2=g_{\mu\nu}dx^\mu dx^\nu=
  -A^2N dt^2 + N^{-1} dr^2 + r^2 (d\theta^2 + \sin^2\theta d\phi^2)
\ , \end{equation}
with
\begin{equation}
N=1-\frac{2m}{r}
\ . \end{equation}
For the gauge and Higgs field we employ the
spherically symmetric ansatz \cite{jul}
\begin{equation}
\vec A_t = \vec e_r J(r) v
\ , \end{equation}
\begin{equation}
\vec A_r=0 \ , \ \ \
\vec A_\theta =  -\vec e_\phi \frac{1- K(r)}{e} \ , \ \ \
\vec A_\phi =   \vec e_\theta \frac{1- K(r)}{e} \sin \theta
\ , \end{equation}
and 
\begin{equation}
\vec \phi = \vec e_r H(r) v
\ , \end{equation}
with unit vectors $\vec e_r$, $\vec e_\theta$ and $\vec e_\phi$.
For $A_t^a=0$, gravitating monopole solutions are obtained
\cite{ewein,bfm,aichel}.

We now introduce the dimensionless coordinate $x$
and the dimensionless mass function $\mu$,
\begin{equation}
x = e v r \ , \ \ \ \mu=e v m
\ , \label{xm} \end{equation}
as well as the coupling constants $\alpha$ and $\beta$,
\begin{equation}
\alpha^2 = 4 \pi G v^2 \ , \ \ \ \beta^2 = \frac{\lambda}{g^2}
\ . \end{equation}

The $tt$ and $rr$ components of the Einstein equations then yield
the equations for the metric functions,
\begin{eqnarray}
\mu'&=&\alpha^2 \Biggl( \frac{x^2 J'^2}{2 A^2} 
   + \frac{J^2 K^2}{A^2 N}
   + N K'^2 + \frac{1}{2} N x^2 H'^2
\nonumber\\
 & &\phantom{ \alpha^2 \Biggl( }
   + \frac{(K^2-1)^2}{2 x^2} + H^2 K^2
   + \frac{\beta^2}{4} x^2 (H^2-1)^2 \Biggr)
\ , \end{eqnarray}
and
\begin{eqnarray}
 A'&=&\alpha^2 x \Biggl(
    \frac{2 J^2 K^2}{ A^2 N^2 x^2}
   + \frac{2 K'^2}{x^2} + H'^2 \Biggr) A
\ , \label{eqa} \end{eqnarray}
where the prime indicates the derivative with respect to $x$.
For the matter functions we obtain the equations
\begin{eqnarray}
(A N K')' = A K \left( \frac{K^2-1}{x^2} + H^2 - \frac{J^2}{A^2 N} 
 \right)
\ , \end{eqnarray}
\begin{equation}
\left( \frac{x^2 J'}{A} \right)' = \frac{2 J K^2}{AN}
\ , \end{equation}
and
\begin{eqnarray}
( x^2 A N H')' = A H \left( 2 K^2 + \beta^2 x^2 (H^2-1) \right)
\ . \end{eqnarray}

A special solution of these equations is the embedded RN solution
with mass $\mu_\infty$,
unit magnetic charge and arbitrary electric charge $Q$,
\begin{equation}
\mu(x) = \mu_\infty - \frac{\alpha^2 (1 + Q^2)}{2x} , \ \ \ A(x)=1 
\ , \end{equation}
\begin{equation}
K(x)=0 \ , \ \ \
J(x) = J_\infty - \frac{Q}{x} \ , \ \ \ 
H(x)=1 
\ . \label{Q} \end{equation}
The corresponding extremal RN solution has horizon $x_{\rm H}$,
\begin{equation}
x_{\rm H} = \mu_\infty  = \alpha \sqrt{1 + Q^2}
\ . \label{RN} \end{equation}

\section{Dyon Solutions}

Let us first consider the globally regular particle-like solutions
of the SU(2) EYMH system.
Requiring asymptotically flat solutions implies
that the metric functions $A$ and $\mu$ both
approach a constant at infinity.
We here adopt
\begin{equation}
A(\infty)=1
\ , \end{equation}
and $\mu(\infty)=\mu_\infty$ represents the dimensionless mass
of the solutions.
The matter functions also approach constants asymptotically,
\begin{equation}
K(\infty)=0 \ , \ \ \ J(\infty)= J_\infty \ , \ \ \ H(\infty) = 1
\ , \end{equation}
where for magnetic monopole solutions $J_\infty = 0$.
The asymptotic fall-off of the function $J(x)$ determines the dimensionless
electric charge $Q$ (see eq.~(\ref{Q})).

Regularity of the solutions at the origin requires 
\begin{equation}
\mu(0)=0
\ , \end{equation}
and \cite{jul}
\begin{equation}
 K(0)=1 \ ,  \ \ \ J(0) = 0 \ , \ \ \ H(0)=0
\ . \label{bc0} \end{equation}

The globally regular dyon solutions have many features in common
with the globally regular monopole solutions.
In the Prasad-Sommerfield limit, $\beta=0$,
the dyon solutions in flat space are known analytically \cite{jul},
whereas for finite $\beta$ they are obtained numerically
\cite{yves}.
In the presence of gravity, the corresponding gravitating dyon
solutions extend up to
a maximal value of the coupling constant $\alpha$.
Beyond this value no dyon solutions exist.
For small values of $\beta$, the fundamental dyon branch 
does not end at the maximal value $\alpha_{\rm max}$.
Instead it exhibits a small spike there and bends backwards,
up to the critical coupling constant $\alpha_c$.
Since variation of the coupling constant $\alpha^2=4\pi G v^2$
can be considered in two different ways,
either as changing $G$ and keeping $v$ fixed, or vice versa,
for small $\beta$ the fundamental dyon branch
can be interpreted as obtained by first varying $G$
up to the maximal value $\alpha_{\rm max}$,
and then varying $v$ up to the critical value $\alpha_c$.

At the critical value $\alpha_c$
the fundamental dyon branch reaches a limiting solution and
bifurcates with the branch of extremal RN solutions of
unit magnetic charge and electric charge $Q$.
The fundamental dyon branch is thus completely analogous 
to the fundamental monopole branch \cite{bfm}.
This is demonstrated for the normalized mass $\mu_\infty/\alpha^2$
of the dyon solutions with electric charge $Q=1$ and $\beta=0$
in Fig.~1, where for comparison also the monopole solutions 
($Q=0$, $\beta=0$) are shown,
together with the extremal RN solutions of
unit magnetic charge and electric charge $Q=1$ as well as $Q=0$.
The normalization in Fig.~1 is chosen to obtain the finite mass
of the flat space solutions in the limit $\alpha \rightarrow 0$.
The ADM mass 
\begin{equation}
m_{\rm ADM}= \frac{4 \pi v}{e} \frac{\mu_\infty}{\alpha^2}
\end{equation}
can be read off the figure.
For the dyon branch we find a critical coupling of $\alpha_c = 1.404$,
while for the monopole branch it is $\alpha_c = 1.386$ 
\cite{bfm}.

Along the fundamental branch 
the dyon functions approach limiting functions,
when $\alpha \rightarrow \alpha_c$.
As for the monopole solutions, the
metric function $N(x)$ of the dyon solutions develops a minimum, 
which decreases monotonically along the fundamental branch.
In the limit $\alpha \rightarrow \alpha_c$, the minimum approaches zero
at $x_c = \alpha_c \sqrt{1+Q^2}$.
The limiting metric function then consists of an inner part, $x \le x_c$,
and an outer part, $x \ge x_c$.
For $x \ge x_c$,
the limiting metric function corresponds to the metric function $N(x)$
of the extremal RN black hole with $\alpha_c$,
unit magnetic charge and electric charge $Q$.
Likewise the other functions approach limiting functions,
when $\alpha \rightarrow \alpha_c$,
which for $x \ge x_c$ correspond to those
of the extremal RN black hole with $\alpha_c$,
unit magnetic charge and electric charge $Q$.

The limit $\alpha \rightarrow \alpha_c$ is demonstrated in Fig.~2 
for the matter function $J(x)$ of the dyon solution with
electric charge $Q=1$ and $\beta=0$.
In the figure the occurrence of the spike is seen,
since the function $J(x)$ does not reach the limiting function
at the maximal value $\alpha_{\rm max}$,
but instead at the critical value $\alpha_c$.
The limiting function is identically zero for $x \le x_c$
and coincides with the RN function for $x \ge x_c$. 
The limiting behaviour of the other functions of the dyon solutions
is analogous to those of the monopole solutions,
shown in \cite{ewein,bfm}.

Beside the branch of fundamental dyon solutions
there are branches of excited dyon solutions.
The gauge field function $K(x)$ of the $n$-th excited dyon solution 
has $n$ nodes,
whereas the gauge field function of the fundamental dyon
solution decreases monotonically to zero.
The excited dyon solutions also exist only below some maximal value 
of the coupling constant $\alpha$.
Since these excited solutions have no flat space counterparts,
the variation of $\alpha$ along a branch of excited solutions
must be interpreted as a variation of $v$ while $G$ is kept fixed.
In the limit $\alpha \rightarrow 0$ the Higgs field vacuum
expectation value therefore vanishes, while $G$ remains finite.
Because of the particular choice of dimensionless variables (\ref{xm}),
in this limit the solutions shrink to zero size
and their mass $\mu$ diverges.
The coordinate transformation $\tilde x = x/\alpha$ \cite{bfm}
leads to finite limiting solutions in the limit $\alpha \rightarrow 0$. 
In fact, the transformed excited monopole solutions 
approach the Bartnik-McKinnon solutions \cite{bfm}, 
and so do the transformed excited dyon solutions.
This is seen in Fig.~3, where
we show the appropriately normalized mass $\mu_\infty/\alpha$ 
as a function of $\alpha$
for the first excited dyon branch
with electric charge $Q=1$ and $\beta=0$.
For comparison also the
first excited monopole branch is shown.

Further details will be given elsewhere \cite{we}.

\section{Black Hole Solutions}

We now turn to the dyonic black hole solutions of the
SU(2) EYMH system.
Imposing again the condition of asymptotic flatness,
the black hole solutions satisfy the same
boundary conditions at infinity 
as the regular solutions.
The existence of a regular event horizon at $x_{\rm H}$
requires
\begin{equation}
\mu(x_{\rm H})= \frac{x_{\rm H}}{2}
\ , \end{equation}
and $A(x_{\rm H}) < \infty $,
and the matter functions must satisfy
\begin{eqnarray}
 {N' K' }|_{x_{\rm H}} =  \left. K \left( \frac{K^2-1}{x^2} + H^2 
 \right) \right|_{x_{\rm H}}
\ , \end{eqnarray}
\begin{equation}
 J |_{x_{\rm H}} =0
\ , \end{equation}
and
\begin{eqnarray}
 {x^2  N' H' }|_{x_{\rm H}} =  \left. 
   H \left( 2 K^2 + \beta^2 x^2 (H^2-1) \right)
  \right|_{x_{\rm H}}
\ . \end{eqnarray}

Again, the SU(2) EYMH ``black holes within dyons''
have many features in common with
the ``black holes within monopoles''.
In particular, for a given coupling constant $\alpha$,
the black hole solutions corresponding to the fundamental dyon branch 
emerge from the globally regular solution
in the limit $x_{\rm H} \rightarrow 0$ and
persist up to a critical maximal value of the horizon radius.
In Fig.~4 we exhibit the mass of
``black holes within dyons'' with electric charge $Q=1$ and $\beta=0$
as a function of the horizon radius
for several values of $\alpha$,
together with the corresponding branches of RN solutions
of unit magnetic charge and electric charge $Q=1$.
For smaller values of $\alpha$, 
the black hole solutions merge into non-extremal RN solutions
at a critical value of the horizon radius.
For larger values of $\alpha$,
the black hole solutions show a critical behaviour
analogous to the globally regular solutions,
and bifurcate with an extremal RN solution.
Here, with increasing horizon radius
the limiting solution is reached
at a critical value of the horizon radius
smaller than the horizon radius of the 
extremal RN black hole with the same $\alpha$,
unit magnetic charge and electric charge $Q$.
This critical behaviour is demonstrated for the function $J(x)$ 
of the dyon black hole solutions with electric charge $Q=1$
and coupling constants $\alpha=1$ and $\beta=0$ in Fig.~5.

Further details will be given elsewhere \cite{we}.

\section{Conclusions}

The globally regular dyon solutions have many features in common
with the globally regular monopole solutions.
Like the fundamental monopole branch, the fundamental
dyon branch starts from the corresponding flat space solution
and extends up to a critical value of the coupling constant
$\alpha$.
At the critical value, both
the fundamental monopole branch and the fundamental dyon branch 
bifurcate with the corresponding branch of extremal RN solutions.
The critical coupling constant depends slightly on the
electric charge $Q$ of the dyons.

Likewise, beside the fundamental dyon branch
there are branches of excited dyon solutions, which
extend up to a maximal value of $\alpha$.
In the limit $\alpha \rightarrow 0$,
the branches of excited dyon solutions 
tend to the corresponding Bartnik-McKinnon solutions,
like their monopole counterparts.

Starting from the stable monopole solutions in flat space,
the solutions on the fundamental monopole branch remain stable
for $\alpha \le \alpha_{\rm max}$
\cite{holl}.
In contrast, the classical dyon solution in flat space is unstable,
since the mass can be lowered continuously, 
by lowering the electric charge,
as long as there is no charge quantization
\cite{jul}.
This indicates, that the gravitating fundamental dyon 
solutions are also unstable.

In analogy to ``black holes within monopoles'' also
``black holes within dyons'' exist.
For a given value of the coupling constant $\alpha$,
the ``black holes within dyons'' 
emerge from the globally regular solutions in the limit
$x_{\rm H} \rightarrow 0$ and persist up to
a maximal value of the horizon radius.
For small $\alpha$ the black hole solutions merge
into the corresponding non-extremal RN solutions
at a critical value of the horizon radius,
whereas for larger $\alpha$ (but $\alpha < \alpha_{\rm max}$)
the black hole solutions bifurcate 
at a critical value of the horizon radius
with the corresponding extremal RN solutions.

The static spherically symmetric ``black holes within monopoles'' 
and ``black holes within dyons'' 
provide counterexamples to the ``no-hair conjecture''.
EYMH theory also possesses counterexamples of a different type,
namely static aspherical black holes,
which represent ``black holes within multimonopoles''
\cite{ewein2}.
Beside static axially symmetric black holes,
whose EYM counterparts have recently been obtained 
non--perturba\-tively
\cite{kk}, there are static black holes
with only discrete symmetries \cite{ewein2}.
It presents a challenge to construct such 
black holes with only crystal symmetries non-perturbatively.

\newpage

\begin{figure}
\centering
\epsfysize=11cm
\mbox{\epsffile{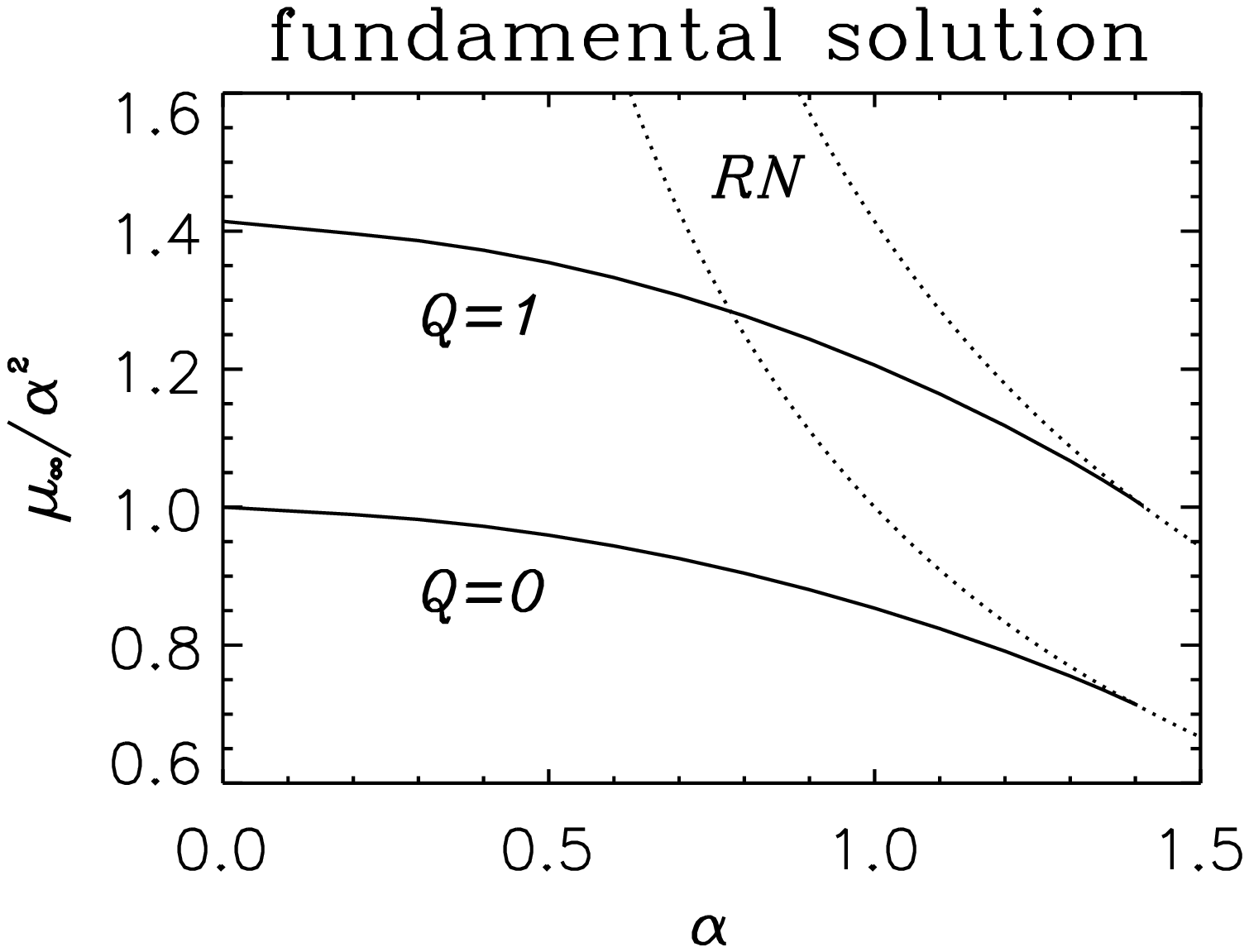}}
\caption{
The mass of the solutions of the fundamental dyon branch for $Q=1$
and the fundamental monopole branch ($Q=0$) 
is shown as a function of the coupling constant $\alpha$
for $\beta=0$ (solid).
Also shown is the mass of the branch of extremal RN solutions with
unit magnetic charge and $Q=1$ as well as $Q=0$ (dotted).
Multiplication with $4\pi v/e $ gives the ADM mass.
At $\alpha=0$ the flat space solutions are obtained.
}
\end{figure}
\newpage

\begin{figure}
\centering
\epsfysize=11cm
\mbox{\epsffile{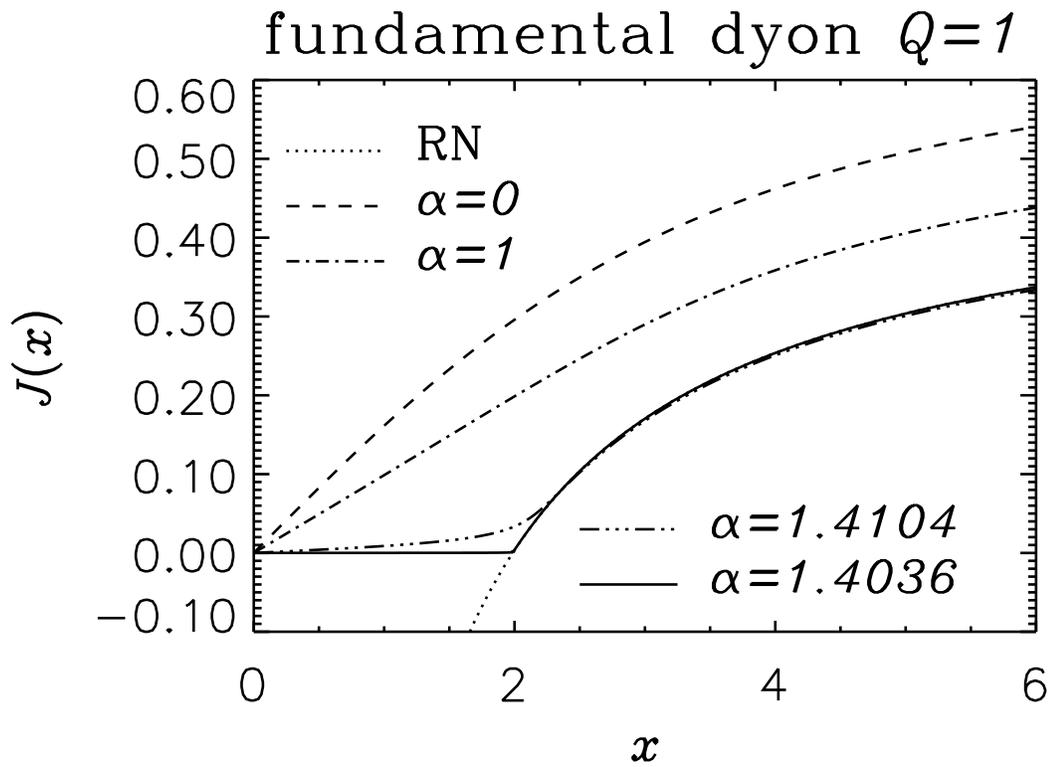}}
\caption{
The function $J(x)$ is shown as a function of $x$
for the regular dyon solution in flat space (dashed)
and the gravitating regular dyon solutions with
$\alpha = 1$ (dot-dashed), $\alpha=1.4104$ (tripledot-dashed)
and $\alpha=1.4036$ (solid) for $\beta=0$.
Also shown is the function $J(x)$ of the corresponding extremal RN solution
with $\alpha=1.4036$ (dotted).
}
\end{figure}
\newpage

\begin{figure}
\centering
\epsfysize=11cm
\mbox{\epsffile{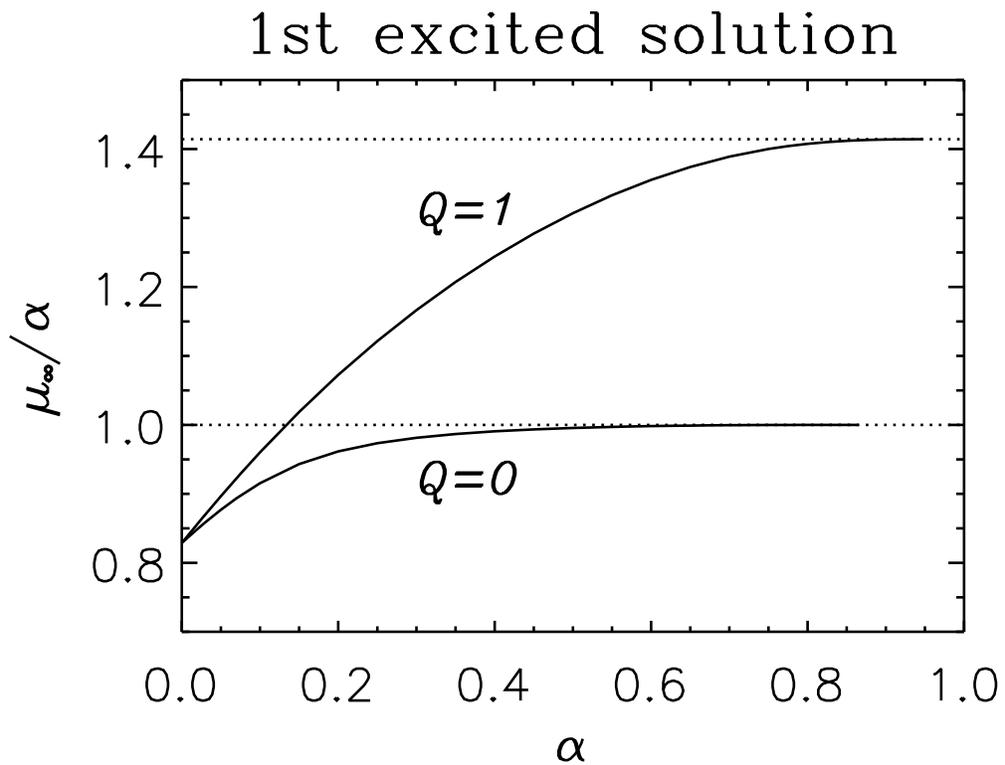}}
\caption{
The mass of the solutions of the 1st excited dyon branch for $Q=1$
and the 1st excited monopole branch ($Q=0$) 
is shown as a function of the coupling constant $\alpha$
for $\beta=0$ (solid).
Also shown is the mass of the branch of extremal RN solutions with
unit magnetic charge and $Q=1$ as well as $Q=0$ (dotted).
The normalization is chosen such that
at $\alpha=0$ the mass of the first Bartnik-McKinnon solution is obtained.
}
\end{figure}
\newpage

\begin{figure}
\centering
\epsfysize=11cm
\mbox{\epsffile{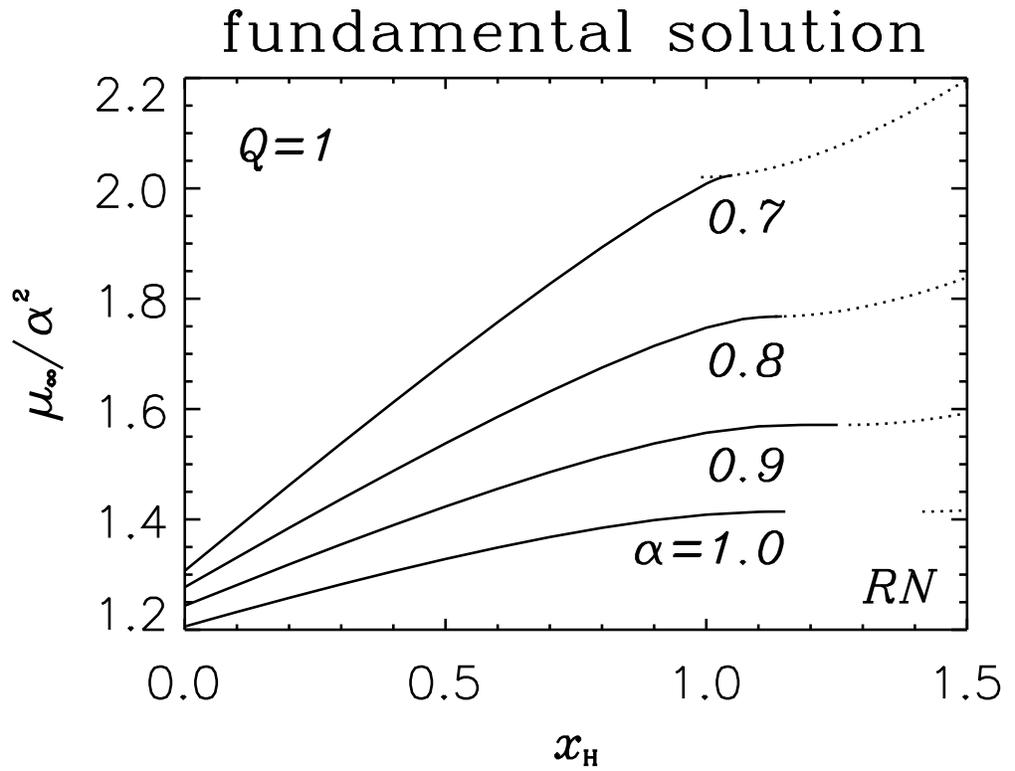}}
\caption{
The mass of the black hole solutions of the fundamental dyon branch
with $Q=1$ is shown as a function of the horizon radius $x_{\rm H}$
for the values of the coupling constant $\alpha=0.7$, 0.8,
0.9 and 1.0 for $\beta=0$ (solid).
Also shown are the masses of the corresponding RN solutions with
unit magnetic charge and $Q=1$ (dotted).}
\end{figure}
\newpage

\begin{figure}
\centering
\epsfysize=11cm
\mbox{\epsffile{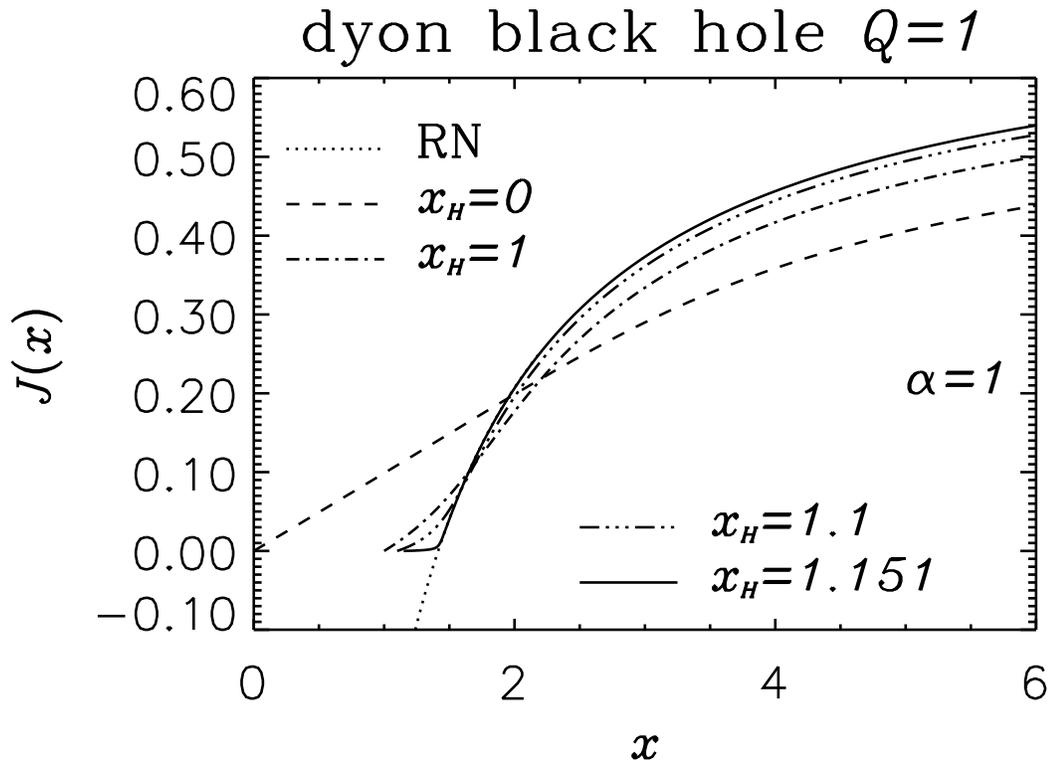}}
\caption{
The function $J(x)$ is shown as a function of $x$
for the gravitating regular dyon solution (dashed),
the dyonic black hole solutions with
$x_{\rm H}=1$ (dot-dashed), $x_{\rm H}=1.1$ (tripledot-dashed)
and $x_{\rm H}=1.151$ (solid) for $Q=1$, $\alpha=1$ and $\beta=0$.
Also shown is the function $J(x)$ of the corresponding extremal RN solution
(dotted).
}
\end{figure}
\end{document}